\documentclass[11pt,a4paper,twoside]{article}
\usepackage{epsfig}
\usepackage{graphics}
\newlength{\figwidth}
\newlength{\figheight}
\def\etal{{et al.}}
\def\smallskip{\\ \indent }

\begin{document}
\setcounter{page}{135}

\title{ THE SM HIGGS-BOSON PRODUCTION IN~\( \gamma \gamma \rightarrow h\rightarrow b\overline{b} \) AT THE PHOTON COLLIDER AT~TESLA  
} 
\author{P.\ NIE\.ZURAWSKI$^1$,
	A.\ F.\ \.ZARNECKI$^1$
	and M.\ KRAWCZYK$^{2,3}$
\\
\\
$^1$ {\it  Institute of Experimental Physics, Warsaw University, Poland}
\\
$^2$ {\it Theory Division, CERN, Switzerland }\\
$^3$ {\it Institute of Theoretical Physics, Warsaw University, Poland }
}
\date{}
\maketitle
\begin{abstract}
 Measuring the \( \Gamma (h\rightarrow \gamma \gamma ){\rm {Br}}
(h\rightarrow b\overline{b}) \) decay
at the photon collider at TESLA is studied for a Standard Model 
Higgs boson of mass \( m_{h}=120 \) GeV. The main background due to the 
process \( \gamma \gamma \rightarrow Q\overline{Q}(g) \), where \( Q=b,\, c \),
is estimated  using the NLO QCD program (G.~Jikia); the results obtained 
are compared with 
the
LO estimate. Using a realistic luminosity spectrum and 
performing a detector simulation, 
 we find that 
\( \Gamma (h\rightarrow \gamma \gamma ){\rm Br}(h\rightarrow b\overline{b}) \)
can be measured with an accuracy better than 2\%
after one year of photon collider running.
\end{abstract}

\begin{figure}[h]
{\centering \resizebox*{!}{\figheight}%
            {\includegraphics{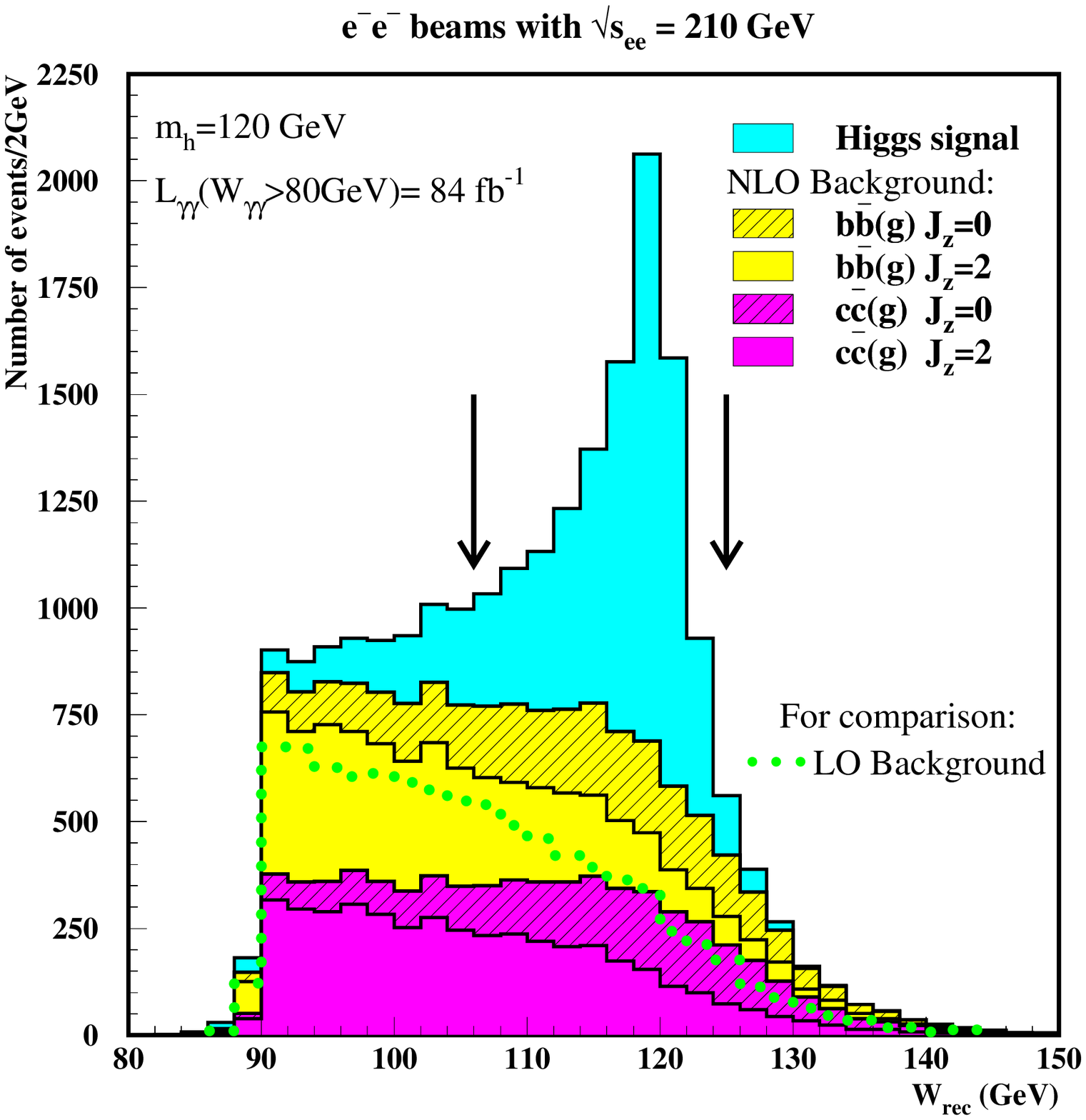} \includegraphics{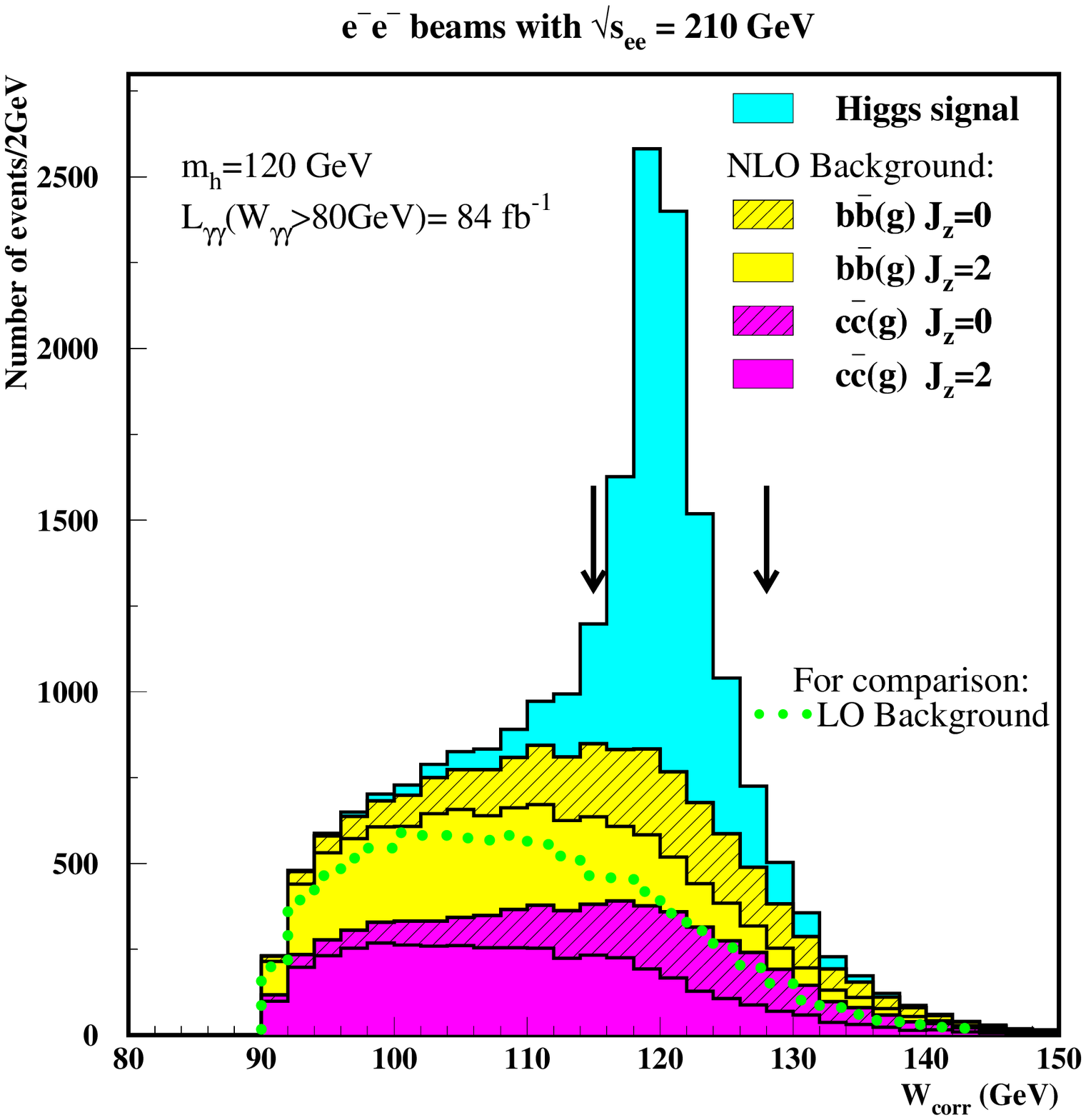}}  \par}

\caption{\label{fig:ResultWithNLOBackgd}
Reconstructed invariant mass \protect\( W_{rec}\protect \) (left) and corrected invariant mass \protect\( W_{corr}\protect \) (right)
distributions for the selected $b\bar{b}$ events.
Contributions of the signal, due to the Higgs boson with a mass 
$m_h = 120$ GeV, and of the heavy-quark 
 background, calculated in the NLO QCD, are indicated. For
 comparison, the LO background estimate  is also plotted (dots). 
Arrows indicate the mass window optimized for the measurement of the 
\( \Gamma (h\rightarrow \gamma \gamma ){\rm Br}(h\rightarrow b\overline{b}) \).
}
\end{figure}

A photon collider option of the TESLA
$e^+e^-$ collider \cite{TDR} offers a  unique possibility to produce
the Higgs boson as an \( s \)-channel resonance and to determine its properties  
with a high accuracy.  The neutral Higgs boson
couples to the photons  through a loop 
with  the massive charged particles,
thus $h\gamma \gamma$ coupling
is sensitive to contributions of new particles.
%
The SM Higgs boson with a  mass below \( \sim 140\) GeV is expected
to decay mainly into the \( b\bar{b} \) final state. 
Here we consider the process \( \gamma \gamma \rightarrow h\rightarrow 
b\overline{b} \) for a Higgs-boson mass of \( m_{h}=120 \) GeV
at a photon collider at TESLA \cite{NZKBB}.
Both the signal and  background
events are generated according to a  realistic photon--photon luminosity
spectrum.
A  simulation
of the detector response is incorporated as well. \smallskip
%
%
%
%
In  the analysis we use the CompAZ parametrization \cite{CompAZ} of 
the spectrum \cite{V.Telnov} to generate  energies of the colliding photons.
For the energy of primary electrons
 \( \sqrt{s_{ee}} =  2 E_{e} \) = 210 GeV,  we obtain a
 peak of the \( J_{z} \) = 0 component 
of the photon--photon luminosity spectrum 
at the invariant mass of the two colliding photons 
\( W_{\gamma \gamma } \) equal to 120 GeV.
%
%
We assume the integrated luminosity
of the primary $e^- e^-$ beams equal to \( L_{ee}^{geom} =502 \; \rm fb^{-1} \)
\cite{V.Telnov}. \smallskip 
%
%
%
A generation of signal events was done with
PYTHIA 6.205 \cite{PYTHIA}, with the parameters for a Higgs boson
as in  HDECAY \cite{HDECAY}. 
A parton shower algorithm, implemented in PYTHIA,
was used to generate the final-state particles. 
The background events due to processes 
\( \gamma \gamma \rightarrow b\bar{b}(g),\, \, c\bar{c}(g) \)
were  generated using the program written by G.~Jikia \cite{JikiaAndSoldner},
where a complete  NLO QCD  calculation for the production of  massive
quarks is performed within the massive-quark scheme.\footnote{%
Other background contributions, from the resolved photon(s)
 interactions and the overlaying events, were found to be negligible.} 
For a comparison we generated also the  LO background events
 as  implemented in  PYTHIA.
The fragmentation was performed using the PYTHIA program. A fast
simulation  for a TESLA detector (SIMDET 3.01 \cite{SIMDET})
was used  to model a detector performance. The jets were reconstructed with the Durham algorithm (\( y_{cut} = 0.02 \)).
The double $b$-tag was required to select the signal
\( h\rightarrow b\bar{b} \) events 
(\( \varepsilon _{bb}=70\% \)
 and 
\( \varepsilon _{cc}=3.5\% \) were assumed). 
The following cuts were used  to 
select reconstructed $b \bar{b}$ events:
(1) a total visible energy \( E_{vis} > 90 \) GeV;
(2) the ratio of the total longitudinal momentum of all 
 observed particles to the total visible energy 
\( |P_{z}|/E_{vis}<0.1 \);
(3) a number of jets \( N_{jets}=2,\, 3 \); 
(4) for each jet
we require 
   \( |\cos \theta _{jet}|<0.75 \).
The obtained
 distributions of the reconstructed \( \gamma \gamma  \) invariant
mass \( W_{rec} \) are shown in 
Fig. \ref{fig:ResultWithNLOBackgd} (left). 
%
%
The expected relative statistical error of
\( \Gamma (h\rightarrow \gamma \gamma ){\rm Br}(h\rightarrow b\overline{b}) \) is equal to
%
%
\( \sqrt{N_{obs}}/(N_{obs}-N_{bkgd}) \).
%
If estimated from the  selected
mass region,
it is equal to 1.9\%. 
We introduce  the corrected, reconstructed invariant mass as: 
%
\(
W_{corr}\equiv \sqrt{W^{2}_{rec}+2P_{T}(E_{vis}+P_{T})}
\).
%
%
The distributions of the \( W_{corr} \) 
are shown in Fig.~\ref{fig:ResultWithNLOBackgd} (right). 
In the selected \( W_{corr} \) region one 
achieves an relative accuracy
%
%
\(
\Delta \left[ 
\Gamma (h\rightarrow \gamma \gamma ){\rm Br}(h\rightarrow b\overline{b})\right] %
/\left[ \Gamma (h\rightarrow \gamma \gamma ){\rm Br}(h\rightarrow b\overline{b})
         \right] =1.7\% \; . \) \smallskip
%
%
%
%
%
%
Assuming \( {\rm Br}(h\rightarrow b\overline{b}) \) 
will be measured to 1.5\% \cite{Brient}, Higgs-boson partial width 
\( \Gamma (h\rightarrow \gamma \gamma ) \) can be extracted with accuracy of 2.3\%.
Using in addition the result from the $e^+ e^-$ Linear Collider for 
\( {\rm Br}(h\rightarrow \gamma \gamma) \) 
\cite{Boos}, one can extract 
\( \Gamma_{\rm tot} \) with a precision of 10\%.
%
%
%
%
%
%
%
%
%
%

\vspace*{-0.8cm}


\end{document}